\documentclass[12pt]{article}
\usepackage{amsfonts,amsmath}
\usepackage{latexsym,graphicx,color,epic}

\newcommand{\be}{\begin{eqnarray}}
\newcommand{\ee}{\end{eqnarray}}

\newcommand{\KE}{K(E_{10})}
\newcommand{\Ke}{K({\mathfrak e}_{10})}
\newcommand{\E}{E_{10}}
\newcommand{\ints}{{\mathbb{Z}}}
\newcommand{\mf}[1]{{\mathfrak{#1}}}
\newcommand{\xo}{{\bf x}_0}
\newcommand{\G}{\Gamma}
\newcommand{\nn}{\nonumber}
\newcommand{\lae}{{\mathfrak{e}}}
\newcommand{\cD}{{\cal{D}}}
\newcommand{\cQ}{{\cal{Q}}}
\newcommand{\p}{\partial}
\newcommand{\eps}{\epsilon}

\begin{document}
{\flushright AEI-2007-015\\[20mm]}

\begin{center}

{\LARGE\sc Unifying R-symmetry in M-theory\footnote{
  \em Based on a talk given at
  the International Congress on Mathematical Physics (ICMP),
  Rio de Janeiro, Brazil, 6 August -- 11 August 2006}}\\[20mm]

{\bf Axel Kleinschmidt}\\[5mm]

{\em Max Planck Institute for Gravitational Physics, Albert
Einstein Institute,\\
Am M\"uhlenberg 1, 14476 Potsdam, Germany}\\[15mm]

\begin{abstract}
\noindent
In this contribution we address the following question: Is there
a group with a fermionic presentation which unifies
all the physical gravitini and dilatini of the maximal supergravity
theories in $D=10$ and $D=11$ (without introducing new degrees of
freedom)? The affirmative answer relies on a new mathematical object
derived from the theory of Kac--Moody algebras, notably $E_{10}$. It
can also be shown that in this way not only the spectrum but also
dynamical aspects of all supergravity theories can be treated
uniformly. 
\end{abstract}

\end{center}

\begin{section}{Introduction}

One of the major themes in string theory has been unification. By
this we mean that hitherto unrelated theories and their properties
are interpreted as different aspects of a single more general
and more fundamental model.
In a very broad sense these advances can be called {\em duality
relations} and typically were first
largely conjectural but were substantiated later by computations.
Among the most far-reaching of these duality conjectures is the
M-theory conjecture \cite{Wi95,HuTo95} which states that all five
known superstring theories have a common origin which is usually
termed M-theory. However, no complete definition
of M-theory is known to date.

It is the aim of this contribution to illustrate how the M-theory
picture can be made more precise by studying a somewhat restricted 
set-up. More precisely, we will focus on
\begin{itemize}
\item{the low energy effective theories with maximal
  supersymmetry. These are the $D=11$ supergravity theory and the
  $D=10$ type IIA and type IIB theories.}
\item{the fermionic sectors of these theories. Since all these
  models have maximal supersymmetry they have the same number of
  physical degrees of freedom, equal to $128$, in their fermionic
  (and bosonic) sectors. However, these are distributed
  differently into representations of the relevant Lorentz and
  R-symmetries.}
\end{itemize}

The fermionic spectra can be summarised by the following table.

\renewcommand{\baselinestretch}{1.3}
\begin{table}[h!]\centering
\begin{tabular}{|l|c|l|}
\hline
Theory & Lorentz \& R-symmetry & Representation\\
\hline\hline
$D=11$ & $SO(1,10)$ & Gravitino $\psi_M$\\
 &&\quad\quad $({\bf 320} \oplus {\bf 32})$\\\hline
$D=10$ IIA & $SO_A(1,9)$ & Two gravitini
  $\psi_\mu^{(1)},\psi_\mu^{(1)}$ (achiral)\\&&
   \quad\quad$({\bf 144}\oplus {\bf \overline{16}})\oplus
    ({\bf \overline{144}}\oplus {\bf 16})$\\
 && Two dilatini $\lambda^{(1)},\lambda^{(2)}$ (achiral)\\&&
   \quad\quad${\bf 16}\oplus {\bf \overline{16}}$\\\hline
$D=10$ IIB & $SO_B(1,9)\times SO(2)$ & Two gravitini
  $\psi_\mu^{(1)},\psi_\mu^{(1)}$ (chiral)\\&&
   \quad\quad$(({\bf 144},{\bf 2})\oplus ({\bf \overline{16}},{\bf 2}))$\\
 && Two dilatini $\lambda^{(1)},\lambda^{(2)}$ (chiral)\\&&
   \quad\quad$({\bf \overline{16}},{\bf 2})$\\\hline
\end{tabular}
\renewcommand{\baselinestretch}{1}
\caption{\label{mreptab}\em Fermionic representations of the
various maximal supergravity theories in $D=10$ and $D=11$.}
\end{table}

In this table, the relevant irreducible representations of the
different Lorentz groups are indicated. Since a gravitino is a
vector-spinor it always consists of a $\G$-traceless part and a
pure $\G$-trace; in the $D=11$ case these are the ${\bf 320}$ and
${\bf 32}$ respectively. As is well known, the type IIA theory
employs spinors of both chiralities of the $D=10$ Lorentz group
whereas in type IIB only one chirality is used. The known
relations for the various Lorentz groups following from
dualities are:
\be\label{mreq}
& SO_B(1,9)&\nn\\
& \cup & \\
\cdots\,\, \subset & SO(1,8) & \subset\,\, SO_A(1,9)\,\, \subset\,\,
SO(1,10)\nn 
\ee
I.e. the type IIA theory is contained in the $D=11$ theory (via
dimensional reduction), but the type IIB theory is not. However,
after reduction to $D=9$ the IIA and IIB theories agree. The
M-theory conjecture now stipulates that there be a unifying
structure to this diagram. This is the first question we address
here: {\em Is there a group $K$ which has subgroups $SO(1,10)$,
$SO_A(1,9)$ and $SO_B(1,9)\times SO(2)$ with embedding relations
given as in (\ref{mreq}) and with a spinor representation which
decomposes under these subgroups into the representations of
table~\ref{mreptab}?} This {\em kinematical}
question will be answered in the
affirmative in section~\ref{kinsec}.

The second question addressed in this contribution is:
{\em Is there a dynamical
equation with explicit $K$ symmetry for the $K$ spinor
representation (constructed in the answer to the first question)
which reduces to the dynamics of the fermionic fields of the
various supergravity theories?} This {\em dynamical} question will
receive a
partially affirmative answer in section~\ref{dynsec}.\\

The work reported on here is based on the papers
\cite{DaKlNi06a,dBHP06,KlNi06a,DaKlNi06b,KlNi06b} which studied
the fermionic sectors of maximal supergravity theories and their
symmetries. The approach taken there (and also here) arises from
known results of unifying symmetries in the corresponding bosonic
sector. In particular, it was shown in
\cite{We01,SchnWe01,SchnWe02,DaHeNi02,We03a,NiFi03,KlSchnWe04}
that the indefinite Kac--Moody algebras $E_{10}$ and $E_{11}$
contain the correct spectra at low levels in so-called level
decompositions. The Dynkin diagram of $E_{10}$ is given in
figure~\ref{e10dynk} and the uncanny resemblance of the right end of
the Dynkin diagram to the structure in (\ref{mreq}) is not
accidental. $E_{11}$ contains the correct fields as 
covariant Lorentz tensor whereas $E_{10}$ breaks Lorentz symmetry
with only manifest spatial Lorentz symmetry.\footnote{For this
  reason the level decomposition of $E_{10}$ does not contain
  anti-symmetric ten-form fields for type IIA and type IIB
  \cite{KlNi04b} whereas $E_{11}$ does \cite{KlSchnWe04}. That
  non-propagating ten-forms, as predicted by $E_{11}$ are
  compatible with the supersymmetry algebra was verified in
  \cite{Beetal05,Beetal06}.}
The bosonic low level spectra correspond to the bosonic version of the
first, kinematical 
question raised above --- in order to address the second,
dynamical question for bosons further `specifications' are required.
For $E_{11}$, West proposed in \cite{We01} that M-theory should be a
non-linear realisation of $E_{11}$; if space-time also carries an
$E_{11}$ structure it nicely incorporates all central charges of the
$D=11$ supersymmetry algebra \cite{We03b} but also infinitely many
more new coordinates. The same $E_{11}$ structure was found for the
bosonic sectors of (massive) type IIA and type IIB in
\cite{We01,SchnWe01,SchnWe02}. 
For $E_{10}$, Damour, Henneaux and Nicolai
proposed in \cite{DaHeNi02} a one-dimensional non-linear
$\sigma$-model based on an $E_{10}$ coset space and demonstrated
that at low levels null geodesic motion on this coset space
is equivalent to the
$D=11$ dynamics around a fixed spatial point truncated roughly
after first spatial gradients. Higher order spatial gradients
were conjectured to arise via the higher levels in the
decomposition. This picture was extended to (massive)
type IIA and type IIB in \cite{KlNi04a,KlNi04b}. A model combining
$E_{11}$ with the null geodesic idea of $E_{10}$ was given in
\cite{EnHo04a,EnHo04b}.

In this contribution we will work with $E_{10}$ because in this
case we can give a more complete answer to the kinematical and
dynamical questions raised above. Since
$E_{10}$ treats time and space asymmetrically, all necessary
requirements for the sought-after `M-theory Lorentz group' $K$
only involve spatial Lorentz groups and their representations.
We will comment on the covariant formulation in the final
section. In order to convey the main ideas we mostly refrain from
introducing intricate notations and outline the logic; more details
can be found in
references~\cite{DaKlNi06a,dBHP06,KlNi06a,DaKlNi06b}. 

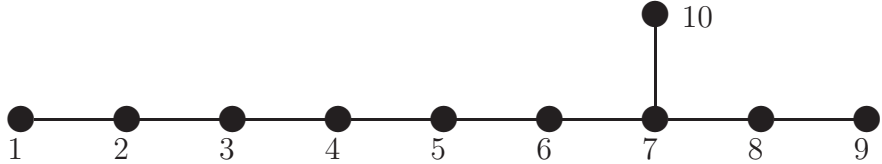
\begin{figure}[t!]
\begin{center}
\scalebox{1}{
\begin{picture}(340,60)
\put(5,-5){$1$} \put(45,-5){$2$} \put(85,-5){$3$}
\put(125,-5){$4$} \put(165,-5){$5$} \put(205,-5){$6$}
\put(245,-5){$7$} \put(285,-5){$8$} \put(325,-5){$9$}
\put(260,45){$10$} \thicklines
\multiput(10,10)(40,0){9}{\circle*{10}}
\multiput(15,10)(40,0){8}{\line(1,0){30}}
\put(250,50){\circle*{10}} \put(250,15){\line(0,1){30}}
\end{picture}}
\caption{\label{e10dynk}\sl Dynkin diagram of $E_{10}$ with numbering
of nodes. $E_{11}$ has an additional node attached with a single line
to node $1$.}
\end{center}
\end{figure}

\end{section}

\begin{section}{Kinematics}
\label{kinsec}

The study of dimensional reduction \cite{CJ79,CrJuLuPo98a} suggests
that the group $K$ we are looking for is $K=K(E_{10})$, the `maximal compact
subgroup' of $E_{10}$. In order to see that this is true we first
need to understand what $\KE$ is. 

\begin{subsection}{Definition of $\lae_{10}$ and $\Ke$}

$K(E_{10})$ is infinite-dimensional and
since global issues are somewhat tricky we will restrict our
attention here to the Lie algebras. The Lie algebra $\Ke$ of $\KE$
is a subalgebra of the Lie algebra $\lae_{10}$ of $\E$. The Lie
algebra $\lae_{10}$ is defined in the Chevalley--Serre
presentation by giving $30$ simple generators
\be
\label{e10gens}
e_i, f_i, h_i \quad\quad (i=1,\ldots,10)
\ee
and their relations (for all $i,j=1,\ldots 10$)\footnote{ad
denotes the adjoint action: $(\text{ad}\, e_i) e_j = [e_i,e_j]$.}
\begin{align}
\label{e10rels}
[h_i,e_j] &= A_{ij} e_j\,,&\quad [h_i,f_j] &= -A_{ij} f_j\,&
[e_i,f_j] &= \delta_{ij}h_i\,,\nn\\
{}[h_i,h_j]&=0\,,&\quad(\text{ad}\,e_i)^{1-A_{ij}} e_j &=0\,,&\quad
(\text{ad}\,f_i)^{1-A_{ij}} f_j &=0\,,&
\end{align}
where $A_{ij}$ is the generalised Cartan matrix which can be read
off from fig.~\ref{e10dynk} as follows: $A_{ii}=2$ for
$i=1,\ldots,10$ and if there is a single link between nodes $i$
and $j$ then $A_{ij}=A_{ji} =-1$ and $A_{ij}=0$ otherwise.
$\lae_{10}$ is defined as the Lie algebra with simple generators
(\ref{e10gens}) and relations (\ref{e10rels}).

On $\lae_{10}$ one can define the Chevalley involution $\theta$
acting by
\be
\theta(e_i) = -f_i\,,\quad \theta(f_i) = -e_i\,,\quad
  \theta(h_i)= - h_i
\ee
on the simple generators. The fixed point set of this involution
defines the `compact subalgebra' $\Ke$:
\be
\Ke = \left\{x\in\lae_{10}\,:\, \theta(x) =x \right\}.
\ee
This subalgebra is called compact because it has definite Killing
norm, generalising the notion of compact algebras in the
finite-dimensional case.

It can be shown \cite{Be89} that $\Ke$ is generated by the simple
generators
\be
\label{ke10gens}
x_i = e_i -f_i \quad\quad (i=1,\ldots,10)
\ee
which are manifestly invariant under $\theta$ and defining
relations of the type
\be
\label{ke10rels}
\sum_{k=0}^{1-A_{ij}} C_{ij}^{(k)} (\text{ad}\, x_i)^k x_j = 0,
\ee
where $C_{ij}^{(k)}$ are constant coefficients and can be computed
from the Cartan matrix. This defines a presentation of $\Ke$ in
terms of generators and relations. For both $\lae_{10}$ and $\Ke$
this type of presentation is the only known presentation. Whereas
$\lae_{10}$ is a Kac--Moody algebra with well-defined structure
theory \cite{Ka90}, $\Ke$ is {\em not} a Kac-Moody algebra \cite{KlNi05}
and its general representation theory is unknown. Nevertheless, the
relations (\ref{ke10rels}) are sufficient to establish the
consistency of any tentative representation as we will see below.
All Lie algebras we consider are over the real numbers, in
particular $\lae_{10}$ is in split form.

\end{subsection}

\begin{subsection}{Level decompositions for $D=11$, IIA and IIB}

A more economical and physical description of the generators of
$\lae_{10}$ can be obtained via a so-called level decomposition
\cite{DaHeNi02,KlSchnWe04}
where one represents
\be
\label{levdec}
\lae_{10} = \sum_{\ell\in\ints} \lae^{(\ell)}_{10}
\ee
as a graded sum of (finite-dimensional reducible) representation
spaces of a chosen regular subalgebra. The subalgebras of interest
are obtained by removing nodes from the $E_{10}$ Dynkin diagram.
The integer $\ell$ represents the level (if several nodes are
removed it consists of a tuple of integers). Regular subalgebras
of $\lae_{10}$ naturally give rise to subalgebras of $\Ke$.

The subalgebras relevant for $D=11$, type IIA and type IIB are
displayed in table~\ref{subtab}. From the table it is evident that
$\Ke$ admits subalgebras of the type required by condition
(\ref{mreq}) and that these satisfy the necessary embedding
conditions. (Recall that the time coordinate is treated separately
for $E_{10}$ whence we are only dealing with the spatial Lorentz
groups here.)

\renewcommand{\baselinestretch}{1.3}
\begin{table}[b!]
\centering
\begin{tabular}{|l|l|c|c|}
\hline
Theory & Dynkin diagram & Subalgebra of $\lae_{10}$ & ... of
$\Ke$\\
\hline\hline
$D=11$ &
\begin{picture}(120,24)
\multiput(5,2)(14,0){9}{\circle*{6}}
\multiput(8,2)(14,0){8}{\line(1,0){8}}
\put(89,16){\circle{6}} \put(89,5){\line(0,1){8}}
\end{picture}
& $\mf{sl}(10)$ & $\mf{so}(10)$\\\hline
$D=10$ IIA&
\begin{picture}(120,24)
\multiput(5,2)(14,0){8}{\circle*{6}}
\put(117,2){\circle{6}}
\multiput(8,2)(14,0){8}{\line(1,0){8}}
\put(89,16){\circle{6}} \put(89,5){\line(0,1){8}}
\end{picture}
& $\mf{sl}_A(9)$ & $\mf{so}_A(9)$\\\hline
$D=10$ IIB&
\begin{picture}(120,24)
\multiput(5,2)(14,0){7}{\circle*{6}}
\put(103,2){\circle{6}}
\put(117,2){\circle*{6}}
\multiput(8,2)(14,0){8}{\line(1,0){8}}
\put(89,16){\circle*{6}} \put(89,5){\line(0,1){8}}
\end{picture}
& $\mf{sl}_B(9)\oplus \mf{sl}(2)$ & $\mf{so}_B(9)\oplus \mf{so}(2)$\\\hline
\end{tabular}
\renewcommand{\baselinestretch}{1}
\caption{\label{subtab}\em The subalgebras relevant for the
various maximal supergravity theories. Empty nodes are to be
deleted.}
\end{table}

We exemplify the result of the level decomposition for the $D=11$
case, that is for the case of the depicted $\mf{sl}(10)$
subalgebra of $\lae_{10}$. At level $\ell=0$ the reducible
representation of $\mf{sl}(10)$ turns out to be $\mf{gl}(10)$ with
generators $K^a{}_b$. Moreover, all higher levels are
representations of $\mf{gl}(10)$. Concretely,
\be
\label{a9spec}
\ell = 0 &:& K^a{}_b\nn\\
\ell = 1 &:& E^{abc} = E^{[abc]}\nn\\
\ell = 2 &:& E^{a_1\ldots a_6} = E^{[a_1\ldots a_6]}\\
\ell = 3 &:& E^{a_0|a_1\ldots a_8} = E^{a_0|[a_1\ldots
a_8]}\quad,\quad E^{[a_0|a_1\ldots a_8]}=0\nn\\
\vdots\quad&&\quad\vdots\nn
\ee
Here, ($a,b=1,\ldots,10$) are $\mf{sl}(10)$ vector indices and
$\ell=1,2,3$ are irreducible representations (accidentally).
These tensors suggest a relation to the bosonic fields of
$D=11$ as follows: $\ell=0$ is related to the spatial part
$e_m{}^a$ of the vielbein, $\ell=1$ is related to the
spatial components of the anti-symmetric three-form gauge
potential, $\ell=2$ is related to the Hodge dual of the three-form
potential and $\ell=3$ is related to the dual of the vielbein.
That this is true in the one-dimensional $E_{10}/K(E_{10})$
$\sigma$-model was shown in \cite{DaHeNi02}.

Our interest here is in $\KE$ and therefore we have to form the
invariant combinations of the generators in (\ref{a9spec}) to
obtain
\be
\label{kegens}
\ell=0 &:& J^{ab} = K^a{}_b +\theta(K^a{}_b) = K^a{}_b-K^b{}_a = J^{[ab]}\nn\\
\ell=1 &:& J^{abc} = E^{abc} + \theta(E^{abc})\nn\\
\ell=2 &:& J^{a_1\ldots a_6} = E^{a_1\ldots a_6}
 + \theta(E^{a_1\ldots a_6})\\
\ell=3 &:& J^{a_0|a_1\ldots a_8} = E^{a_0|a_1\ldots a_8}
  + \theta(E^{a_0|a_1\ldots a_8}) \nn\\
\vdots\quad&&\quad\vdots\nn
\ee
For $\Ke$ the level $\ell$ has to be taken with a grain of salt
since it does no longer define a grading but only a filtered
structure. Indeed, examples of $\Ke$ commutation relations are
\cite{DaKlNi06a}
\be
\label{so10rels}
[J^{ab}, J^{cd}] &=&  \delta^{bc} J^{ad} - \delta^{bd} J^{ac}
   -\delta^{ac}J^{bd} + \delta^{ad}J^{bc},\nn\\
{}[J^{a_1a_2a_3},J^{a_4a_5a_6}] &=& J^{a_1\ldots a_6}
  -18 \delta^{[a_1a_2}_{[a_4a_5} J^{a_3]}{}_{a_6]}.
\ee
We see that the first line is the $\mf{so}(10)$ subalgebra of
$\Ke$ and the second line gives generators of `levels' $\ell=2$
and $\ell=0$ on the right hand side in accordance with the
filtered structure. The $\mf{so}(10)$ subalgebra introduces the
invariant $\delta^{ab}$ which can be used to raise and lower the
tensor indices. There are infinitely many more relations than
(\ref{so10rels})
involving all the other infinitely many generators and no closed
form is known for them.

\end{subsection}

\begin{subsection}{Representations of $\Ke$}

By virtue of the presentation of $\Ke$ in terms of generators and
relations in (\ref{ke10gens}) and (\ref{ke10rels}) it is
sufficient to verify a finite number of relations on a tentative
representation. Using the level decomposition one can further
reduce this number by starting from a representation of the
subalgebra (which obviously constitutes a necessary condition). Then
the sufficient consistency conditions involve only levels $\ell=0$ and 
$\ell=1$ (basically since there a only single lines in the
$E_{10}$ Dynkin diagram). For the $\mf{so}(10)$ subalgebra of
$\Ke$ the recipe for constructing $\Ke$ representations is:
\begin{enumerate}
\item{ Start from an $\mf{so}(10)$ representation which we call
  the tentative $\Ke$ representation. This defines
  the action of the $J^{ab}$ generators within $\Ke$ on the
  tentative representation.}
\item{ Make a general ansatz for the action of $J^{abc}$ on the
  tentative representation from $\mf{so}(10)$ representation
  theory.}
\item{Verify that the second line of (\ref{so10rels}) holds for the
  case when some of indices are identical on the tentative
  representation. When some indices are identical the term with six
  anti-symmetric indices drops out.\footnote{That this is sufficient follows
  from the precise expressions for the simple $\Ke$ generators $x_i$
  of (\ref{ke10gens}) in terms of components of the $J^{ab}$ and
  $J^{abc}$ which can be found in \cite{DaKlNi06b}.} If there is a
  solution for the general ansatz then the tentative representation
  gives rise to a full consistent representation of $\Ke$.}
\end{enumerate}
For the other subalgebras the procedures are similar. Since it
involves the tensors arising in the corresponding level
decompositions we do not detail them here in order to keep the
exposition simple.

We now construct the gravitino (vector-spinor)
representation of $\Ke$ following the steps above. The
vector-spinor of $\mf{so}(10)$ is reducible of dimension $320$ and
consists of the irreducible pieces ${\bf 288}\oplus {\bf 32}$
corresponding to the $\G$-traceless part and the $\G$-trace. We
denote the vector-spinor by $\psi_a$ and suppress the spinor
index. The $\mf{so}(10)$ generators $J^{ab}$ act on $\psi_a$
by\footnote{Here, $\G^a$ are the real $(32\times 32)$ $SO(10)$
  $\G$-matrices and $\G^{ab}= \G^{[a}\G^{b]}$ etc.}
\be
\label{lev0}
J^{ab} \psi_c = \frac12\G^{ab}\psi_c + 2 \delta_c^{[a}\psi^{b]}.
\ee
In the general ansatz for the $J^{abc}$ action there are three
terms and the solution to the necessary commutation condition
(\ref{so10rels}) leads to \cite{DaKlNi06a,dBHP06}
\be
\label{lev1}
J^{abc} \psi_d = \frac12\G^{abc}\psi_d +4 \delta_d^{[a}\G^b\psi^{c]}
  - \G_d{}^{[ab} \psi^{c]}.
\ee
That there exists a solution to the consistency condition implies
that there is a representation of $\Ke$ of dimension $320$. One can
check that this is in fact an irreducible representation since the
$\G$-trace no longer separates once $J^{abc}$ is considered. {\em We
have thus proved that $\Ke$ has an irreducible ${\bf 320}$
representation}. Under the $\mf{so}(10)$ subalgebra
it decomposes according to 
\be
{\bf 320} &\longrightarrow& {\bf 288} \oplus {\bf 32}\nn\\
\Ke &\supset & \mf{so}(10)
\ee
as required. We denote this representation by $\Psi$ since it can be
defined independently of the $\mf{so}(10)$ subalgebra under which
it is more conveniently written as $\psi_a$.

We now turn to the decompositions under the subalgebras
relevant for type IIA and type IIB. They were derived in
\cite{KlNi06a} and we reproduce the results here as\footnote{In
  \cite{KlNi06a} the subalgebra $\mf{so}(9,9)$ was chosen for type
  IIA (instead of $\mf{sl}_A(9)$) since this more naturally
  includes the mass term of the massive extension of IIA. That the
  result given here is also correct follows immediately from
  $\mf{so}_A(9)\subset \mf{so}(10)$ and the branching rules for
  these groups.}
\be
\label{iiadec}
{\bf 320} &\longrightarrow& ({\bf 128} \oplus {\bf 16})
\oplus ({\bf 128} \oplus {\bf 16}) \oplus {\bf 16}\oplus {\bf 16}\nn\\
\Ke &\supset & \mf{so}_A(9)
\ee
for type IIA, where the last two ${\bf 16}$s are the dilatini, and
\be
\label{iibdec}
{\bf 320} &\longrightarrow& (({\bf 128},{\bf 2}) \oplus ({\bf 16},{\bf
2}))
  \oplus ({\bf 16},{\bf 2})\nn\\
\Ke &\supset & \mf{so}_B(9)\oplus \mf{so}(2)
\ee
for type IIB. Here, the last doublet of ${\bf 16}$s corresponds to
the IIB dilatini. Since we are only dealing with the spatial
Lorentz group $\mf{so}(9)$ different chiralities are not
properly distinguished. The calculation shows, however, that the
two doublets of ${\bf 16}$s arise differently and in the covariant
calculation one can show that indeed all chiralities also fulfil
the necessary requirements to answer the first question raised in
the introduction affirmatively: The group $\KE$ contains the
subalgebras required by the M-theory picture and has a spinorial
representation with the correct number of components
which branches correctly to the fermionic fields of the maximal
supergravity theories.

\end{subsection}

\end{section}

\begin{section}{Dynamics}
\label{dynsec}

To further substantiate the significance of $\KE$ and its
${\bf 320}$ representation $\Psi$ for an algebraic approach to M-theory
we now turn to studying a dynamical equation for $\Psi$
and its relation to the fermionic dynamics in the
various maximal supergravities.

Since time is treated separately in the $E_{10}$ context and all
dynamical equations in the bosonic sector are time evolution
equations a natural ansatz for the fermionic equation is
\be\label{kespineq}
\cD_t \Psi = 0.
\ee
This is a Dirac equation for the $\Ke$ vector-spinor coupled minimally
to a $\Ke$ connection $\cQ_t$ via the covariant derivative 
\be
\cD_t = \p_t -\cQ_t,
\ee
where $\cQ_t\in \Ke$ acts on $\Psi$ in the ${\bf 320}$
representation. The gauge field $\cQ_t$ transforms under $t$-dependent
local $\KE$ gauge transformations. As an $\Ke$ element,
$\cQ_t$ can be expanded over $\mf{so}(10)$  in the
generators (\ref{kegens}) via
\be
\cQ_t &=& \frac12 Q^{(0)}_{ab} J^{ab} + \frac1{3!}Q^{(1)}_{abc}
  J^{abc}
+ \frac1{6!}Q^{(2)}_{a_1\ldots a_6} J^{a_1 \dots a_6}\nn\\
&&\quad\quad+ \frac1{9!}Q^{(3)}_{a_0|a_1\ldots a_8} J^{a_0|a_1\ldots
  a_8} +\ldots.
\ee
Since the action of all the $\Ke$ generators can be computed
from multiple commutators of (\ref{lev0}) and (\ref{lev1}) the
Dirac equation (\ref{kespineq}) can be evaluated to arbitrary
level. In \cite{DaKlNi06a} it was evaluated up to $\mf{sl}(10)$ level
three which is the level to which the field content (\ref{a9spec}) is 
understood~\cite{DaHeNi02}. The resulting expression contains the
gauge field components $Q^{(\ell)}$ (for $\ell=0,\ldots,3$) contracted
with various $\G$-matrices multiplying the $\mf{so}(10)$ decomposed
vector-spinor $\Psi=(\psi_a)$. Explicitly, we find
\be\label{ke10eq}
\cD_t\psi_c &=& \p_t \psi_c -\frac14 Q_{ab}^{(0)}\G^{ab}\psi_c
  -Q_{ca}^{(0)}\psi^a  
  -\frac1{12}Q_{a_1a_2a_3}^{(1)}\G^{a_1a_2a_3}\psi_c\nn\\
&&  -\frac23Q_{ca_1a_2}^{(1)}\G^{a_1}\psi^{a_2}
    +\frac16Q_{a_1a_2a_3}^{(1)}\G_c{}^{a_1a_2}\psi^{a_3} 
  -\frac1{1440}Q_{a_1\ldots a_6}^{(2)}\G^{a_1\ldots a_6}\psi_c \nn\\
&&  +\frac1{72}Q_{ca_1\ldots a_5}^{(2)}\G^{a_1\ldots a_4}\psi^{a_5}
  -\frac1{180}Q_{a_1\ldots a_6}^{(2)}\G_c{}^{a_1\ldots a_5}\psi^{a_6}\\
&&  -\frac2{3\cdot 8!}Q_{a_0|a_1\ldots a_8}^{(3)}\G_c{}^{a_1\ldots a_8}\psi^{a_0}
  -\frac2{3\cdot 7!}Q_{c|a_1\ldots a_8}^{(3)}\G^{a_1\ldots a_7}\psi^{a_8}\nn\\
&&  -\frac4{3\cdot 8!}Q_{b|ba_1\ldots a_7}^{(3)}\G^{a_1\ldots a_7}\psi_c
  -\frac1{3\cdot 6!}Q_{b|ba_1\ldots a_7}^{(3)}\G_c{}^{a_1\ldots a_6}\psi^{a_7}+\ldots\nn.
\ee
This equation has to be compared with the dynamical equation for the
gravitino in $D=11$ supergravity. From the analysis of the bosonic
sector it is to be expected that gauge-fixing is required in order to
establish a connection between the $\Ke$ equation (\ref{ke10eq}) and
the supergravity equation~\cite{KlNi05}. Indeed it turns out
\cite{DaKlNi06a} that one has to fix a supersymmetry gauge ($\psi_0
-\G_0\G^a\psi_a=0$) for the fermions, a pseudo-Gaussian gauge
($E_t{}^a=0$) for the vielbein and a Coulomb gauge ($A_{0ab}=0$) for
the gauge potential. In this case the supergravity equation (to lowest
fermion order) takes almost the same form as (\ref{ke10eq}) but where the
gauge field components take the values~\cite{DaKlNi06a,DaKlNi06b}
\begin{align}
\label{qdict}
Q^{(0)}_{ab}(t) &= -N\omega_{0\,ab}(t,\xo)\,,&\quad 
Q^{(2)}_{a_1\ldots a_6}(t) &= -\frac{1}{4!} N\eps_{a_1\ldots a_6b_1\ldots
    b_4} F_{b_1\ldots b_4}(t,\xo)\,,&\\
Q^{(1)}_{abc}(t) &= N F_{0abc}(t,\xo)\,,& \quad 
  Q^{(3)}_{a_0|a_1\ldots a_8}(t) &= \frac{3}2 N\eps_{a_1\ldots
  a_8b_1b_2}\omega_{b_1\,b_2 a_0}(t,\xo)\,.&\nn
\end{align}
Here, $\omega_{0\,ab}$ and $\omega_{a\,bc}$ are `electric' and
`magnetic' components of the spin connection in flat indices;
similarly $F_{0abc}$ and $F_{b_1\ldots b_4}$ are electric and magnetic
components of the four-form field strength in flat indices. The lapse
$N=E_t{}^0$ is needed to convert the objects on the right hand sides
into components of a world-line tensor $\cQ_t$. 

The equations (\ref{qdict}) are valid only at a fixed spatial point
$\xo$ and in order to match (\ref{ke10eq}) to the supergravity
equation higher spatial gradients of the fields (and the lapse) have
to be ignored. Furthermore, the spatial spin connection must have
vanishing trace $\omega_{b\,ba}=0$ at $\xo$. More details can be found
in \cite{DaKlNi06a,DaKlNi06b}. 

To summarise, with the use of the `dictionary' (\ref{qdict}) we have
succeeded in turning a truncated version of the $D=11$ gravitino
equation of motion into a $\KE$ covariant Dirac-equation of the type
(\ref{kespineq}). Although not explicitly proved in the type IIB case,
one can expect that the very same equation (\ref{kespineq}) also
describes the correct fermionic dynamics of type IIA and IIB by using
the decompositions of $\Ke$ detailed in (\ref{iiadec}) and
(\ref{iibdec}).  

Thus we arrive at the main result: 
$\KE$ is not only a viable candidate for a kinematical unification of 
the fermionic symmetries of {\em all} maximal supergravity theories
but also can partially be established as a symmetry of the dynamical
equations for the fermions.

\end{section}

\begin{section}{Discussion}

\begin{subsection}{Remarks}

Here, we only briefly sketch some related points and comment on a
fully covariant reformulation of the above results.

There exists also a ${\bf 32}$ representation of $\Ke$ which was
  called `Dirac-spinor' in \cite{dBHP05a,DaKlNi06a}. This
  representation is relevant for the supersymmetry parameter
  $\epsilon$ and similarly has the correct branching to the various
  maximal supergravity theories' Lorentz and R-symmetries
  \cite{KlNi06a}. 

Both the ${\bf 320}$ and the ${\bf 32}$ representations of $\Ke$
  are {\em unfaithful} since they are finite-dimensional
  representations of an infinite-dimensional algebra. This implies
  that $\Ke$ is not a simple Lie algebra but has non-trivial
  quotients. These one arrives at by factoring out the ideals
  associated with the unfaithful representations \cite{DaKlNi06b}. In
  the case of the ${\bf 32}$ Dirac-spinor the quotient is
  $\mf{so}(32)$ which has been conjectured as a {\em generalised
  holonomy} in \cite{DuLiu03,Hu03}. Since $\KE$ acts not only on the
  Dirac-spinor but also on the ${\bf 320}$ gravitino (which $SO(32)$
  does not) it is more general than these conjectured
  holonomies. Furthermore, certain global issues \cite{Ke04} are
  resolved in $\KE$~\cite{DaKlNi06a}.

As mentioned in the introduction, the M-theoretic properties of $\KE$
  were derived following similar results in the bosonic
  sector~\cite{We01,SchnWe01,SchnWe02,DaHeNi02,KlNi04a,KlNi04b}. The
  bosonic fields are realised via a coset construction
  $E_{10}/K(E_{10})$ where $\KE$ also acts as a local gauge
  symmetry. It is non-trivial, but true, that the relation between the
  gauge connection appearing in the bosonic analysis and the one in the
  fermionic analysis are related in precisely the same way to the
  supergravity quantities via (\ref{qdict}).

It can also be shown that $K(E_{11})$ (if equipped with the
  temporal involution of \cite{EnHo04a}) allows for a fermionic
  representation of dimension $352=320+32$ if written over
  $SO(1,10)$ \cite{KlNi06a}. The IIA and IIB decompositions of this
  fully covariant gravitino give the correct achiral and chiral
  fermionic spectra in a covariant fashion so that all the results of
  section~\ref{kinsec} carry over to $K(E_{11})$. However, it is not
  clear how to write a $K(E_{11})$ covariant and space-time covariant
  dynamical equation for this gravitino which generalises the Dirac
  equation (\ref{kespineq}). An obvious candidate is
  \be\label{ke11eq}
  {\cal D}\!\!\!\!/\,\, \Psi =0,
  \ee
  where ${\cal D}\!\!\!\!/= \G^M \cD_M = \G^M(\p_M -
  \cQ_M)$. There are a number of subtleties with this suggestive 
  notation that need to be clarified. Firstly, $\cD_M$ should be
  $K(E_{11})$ covariant meaning that the gauge fields transform
  correctly under $K(E_{11})$. By augmenting an $E_{11}/K(E_{11})$
  coset construction by a Borisov--Ogievetsky type construction as in
  \cite{We01} this 
  can probably be achieved. The second problematic point is the symbol
  $\G^M$ used above since $\G^M$  is not an $K(E_{11})$ invariant  
  tensor and so spoils the $K(E_{11})$ covariance of the equation
  even if $\cD_M$ 
  transforms correctly. In line with the philosophy of \cite{We03b} one
  should probably replace eleven-dimensional space-time indices $M$
  by indices taking  values in an infinite-dimensional highest
  weight representation of  $E_{11}$ generalising the translation
  vector to an $E_{11}$ object. It remains to be seen whether one can
  make sense of (\ref{ke11eq}) in this framework.

\end{subsection}

\begin{subsection}{Outlook}

{}From the discussion in the introduction it is clear that in order
to complete the M-theory picture a number of things need to be
included in the present algebraic framework, the most pressing of
which we briefly discuss now.

Firstly, M-theory should also include the non-maximal
heterotic $E_8\times E_8$ and $SO(32)$ string theories
as well as the $SO(32)$ type I superstring. At low energies this
requires fitting the heterotic $D=10$ supergravity with gauge
groups $SO(32)$ and $E_8\times E_8$ into the $E_{10}$ $\sigma$-model
or some more general model. As a first step it was shown in
\cite{KlNi04a,HiKl06} that the pure type I supergravity (without
any vector multiplets) can be interpreted as a subsector of the
$E_{10}$ model. It would be gratifying to see a relation between the
algebraic approach taken here and the issue of anomaly freedom.

Secondly, M-theory presumably is a theory of strings and other
extended objects. The analysis so far only covered point
particles since properties of the low energy field
theories were studied. It is not clear if the symmetries found
need to be modified when extended objects are also considered.
Results from U-duality \cite{HuTo95,ObPi99} suggest that the
continuous symmetry gets broken to some discrete arithmetic group
and first ideas in this direction were discussed in
\cite{BrGaHe04}. A different route was taken in \cite{DaNi05,DaHaHeKlNi06}
where string induced higher derivative corrections to the low
energy effective action were studied in relation to $E_{10}$ and
good agreement between the algebraic structure and conjectured
properties of these correction terms was found (see also
\cite{LaWe06}). 

Thirdly, the bosonic fields appear through the
infinite-dimensional coset space $E_{10}/K(E_{10})$ whereas the
fermionic fields presently are 
confined to a finite-dimensional, unfaithful representation of
$\KE$. This seems problematic from a supersymmetry point of view. This
dichotomy is partly related to the difference in order of the
equations of motion for bosons and fermions. The fermionic field
equations are first order whereas the bosonic ones are second order
(allowing for dualisations and triggering for example the infinite
duality cascade in $D=2$). It would be nice to overcome this obstacle
through the construction of an appropriate faithful fermionic
representation of $\KE$. 

Finally, on the purely mathematical side it could be hoped that a
proper understanding of the relation between $E_{10}$ and M-theory
may lead to a new presentation of the $E_{10}$ structure itself.
Since its inception in the late 1960s \cite{Ka67,Mo67} the
theory of indefinite Kac--Moody algebras has produced few results
which truly penetrate the structure of these fascinating objects.

\end{subsection}

\end{section}

\vspace{5mm}
{\bf Acknowledgements}

\noindent
The author is grateful to Thibault Damour and Hermann Nicolai for very
stimulating collaborations leading to the results presented here and
would like to thank the organisers of the ICMP 2006 conference for
providing an interesting congress. This work was partially supported
by the European Research and Training Networks `Superstrings'
(contract number MRTN-CT-2004-512194) and `Forces Universe' (contract
number MRTN-CT-2004-005104).

\end{document}